\documentclass{article}
\usepackage{hiph-preprint}
\usepackage{graphicx}
\usepackage{amssymb}
\usepackage{psfig}
\usepackage{epsfig}
\volnumber{22} \issuenumber{1} \edyear{2005}                             
\frompage{000} \topage{000}                                              
\recrevdate{1 January 2005}                                              
\def\rightmark{Energy Systematics: 200 vs. 62.4 AGeV}
\def\leftmark{A. Adil and M. Gyulassy}
 
\title{Systematics of Jet Tomography at RHIC: \\
 $\sqrt{s}=62.4$ vs. 200 AGeV} 
\authors{ 
{A. Adil$^1$ and M. Gyulassy$^{2}$ %
\index{One, A.} 
\index{Two, A.} 
}\\[2.812mm]
{\normalsize
\hspace*{-8pt}$^{1,2}$  Columbia University,\\ 
538 West 120th Street, New York, NY, USA.\\[0.2ex] 
}}
 
\abstract{The collision energy dependence of jet tomography is investigated
within the GLV formalism. We estimate
systematic uncertainties resulting from the interplay of energy
loss fluctuations and the rapid increase of the parton transverse
momentum slopes as  $\sqrt{s}$ decreases from 200 to 62.4 AGeV.}


 
\makeindex
\begin{document}
 
\maketitle

\section{Introduction}\label{intro}
We study the energy systematics of 
jet tomography in nuclear collisions
within the GLV formalism\cite{Gyulassy:2003mc,Gyulassy:2001nm}
in the $\sqrt{s}=$ 62.4 to 200 AGeV range.  Jet tomography at $\sqrt{s}=$ 62.4 AGeV data \cite{Back:2004ra} tests the predicted $\sqrt{s}$ decrease of the QGP density
and the predicted variation of the gluon/quark jet source.
In this paper, we calculate the nuclear modification factor,
$R_{AA}(p_{T}, y=0, \sqrt{s})$, for central $Au-Au$
collisions at both $\sqrt{s}=62.4,\; 200$ AGeV for neutral pions.
Previous predictions for 62 AGeV have been published by Wang\cite{Wang:2003aw}
and Vitev\cite{Vitev:2004gn}.

We concentrate here on the role of energy loss fluctuations
\cite{Gyulassy:2001nm,Wiedemann:2003} on the predicted single
hadron attenuation pattern in order to
gain an estimate of some of the systematic theoretical errors in the
jet tomographic technique.  To isolate
the role of fluctuations we neglect $k_T$ smearing, the Cronin
enhancement, gluon and quark shadowing, and nonperturbative baryon
dynamical contributions that strongly distort the hadron spectra below
$p_{T}< 4-5$ GeV\cite{Gyulassy:2003mc}.  So our results are valid for $\pi^{0}$ spectra for $p_{T}>4-5$ GeV.

We test
the influence of the shape of the energy loss fraction
spectrum, $P(\epsilon, \bar{\epsilon})$
about the mean energy loss fraction, $\bar{\epsilon}$, which was first pointed out as important in \cite{Baier:2001yt}.  This sensitivity to the shape increases as the high $p_{T}$ slopes increase at lower $\sqrt{s}$.
 
\section{Calculation of Spectra and $R_{AA}$}\label{spectra}  

The neutral pion cross section in $pp$ can be calculated using collinear factorized pQCD.  In a dense QCD medium the induced radiative energy loss reduces the initial $p_{T}$ of the jet parton by a fraction $\epsilon$ before hadronization.  In this paper we calculate the $\pi^{0}$ inclusive spectrum as the following.
\begin{equation}
\label{fullaa2} E_{h}\frac{d\sigma_{\pi^0}(\bar{\epsilon})}{d^3p} =K
        \sum_{abcd}
        \int\!\!dx_1 dx_2 d\epsilon  
f_{a/A}(x_1,Q^2) f_{b/A}(x_2,Q^2) \frac{d\sigma^{ab \rightarrow
cd}}{d{\hat t}}
  P(\epsilon,\bar{\epsilon})
\frac{z^*_c}{z_c}
   \frac{D_{\pi^0/c}(z^*_c,Q^2)}{\pi z_c} \,\,\, ,
\end{equation}
where $z^{*}_{c}=\frac{z_{c}}{1-\epsilon}$The inclusive number distribution is calculated by mutiplying this invariant distribution with the the Glauber geometric binary collisions factor, $T_{AA}(b)$.  The functions $f_{a/A}$ and $D_{h/c}$ are the conventional MRS D- distribution function and KKP fragmentation function, respectively.  Gluon number fluctuations are taken into account using the distribution $P(\epsilon,\bar{\epsilon})$, where $\bar{\epsilon}$ is interpreted as the average fractional energy loss and is proportional the local gluon rapidity density.
 We explore two simplified forms of fluctuation distributions to assess some of the systematic uncertainties in the predicted quenching.  One is a ``uniform" model that essentially reproduces the truncated Poisson of \cite{Gyulassy:2001nm,Vitev:2004gn}.  The second is called ''squeezed" because it accumulates strength near the $\epsilon\approx 1$ opaque limit.  This distribution is considered to take into account the alternative branching form of implementing gluon fluctuations.
 
The ``uniform" distribution takes the form,
\begin{eqnarray}
\label{unif} &&P(\epsilon,\overline{\epsilon})=\left\{
\begin{array}{ll}
\frac{\theta(0<\epsilon<2\bar{\epsilon})}{2\bar{\epsilon}}
& {\rm if }\;\; 0<\bar{\epsilon}<0.5 \\
1 & {\rm if }\;\; 0.5<\bar{\epsilon}
\end{array} \right.
\end{eqnarray}
while the ``squeezed" distributon is the following.
\begin{eqnarray}
\label{squez} &&P(\epsilon,\overline{\epsilon})=\left\{
\begin{array}{ll}
\frac{\theta(0<\epsilon<2\bar{\epsilon})}{2\bar{\epsilon}} & {\rm
if }\;\; 0<\bar{\epsilon}<0.5 \\
\frac{\theta(2\bar{\epsilon}-1<\epsilon<1)}{2(1-\bar{\epsilon})}
&{\rm if}\;\; 0.5<\bar{\epsilon}
\end{array} \right.
\end{eqnarray}
The factors of $K$ and $Q^{2}$ are fit to $pp$ data at the requisite COM energies.  $\bar{\epsilon}$ evolves with $\sqrt{s}$ according to the the multiplicity evolution.  We assume that $\bar{\epsilon}_c(\sqrt{s})=\frac{C_c}{C_g}\left(
\frac{{dN_{g}(\sqrt{s})}/{dy}}{{dN_{g}(200)}/{dy}}\right) \bar{\epsilon}_g(200)
$ where $c$ is the parton type and $C_{c/g}$ are the QCD Casimirs.  Thus, the free parameter is $\bar{\epsilon}_{g}(200)$ which is set to fit PHENIX \cite{Adler:2003qi} $\pi^{0}$ data at 200 AGeV.

 Once the spectra for the $A-A$ and $p-p$ reactions at the requisite energies have been calculated, the nuclear modification factor $R_{AA}$ is just the ratio between them.  This gives a range of values for each distribution (``uniform" and ``squeezed") determined by the errors of the data.  For the ``squeezed" distribution, the determined range is $0.65<\bar{\epsilon}_{g}(200)<0.76$ while the ``uniform" distribution can be fit to $R_{AA}$ data with the range $0.70<\bar{\epsilon}_{g}(200)<0.80$.  The ``squeezed" distribution needs a lower average opacity as the distribution itself is biased towards $\epsilon\rightarrow 1$ when $\bar{\epsilon}>0.5$.
 
\begin{figure}[htb]
\epsfig{file=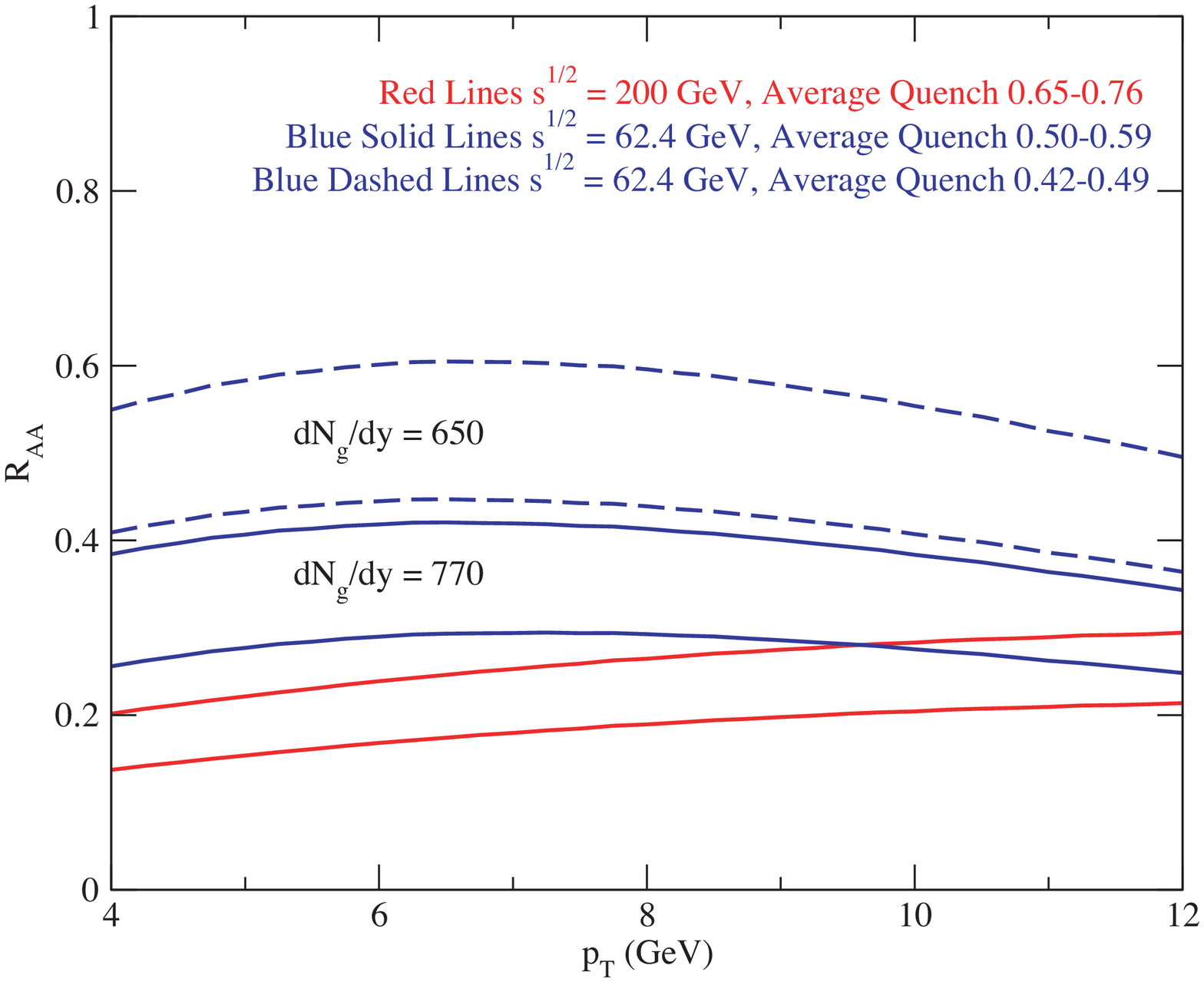,width=2.5in,angle=0}
\hspace{0.1in}
\epsfig{file=raa_pred_unif_624.eps,width=2.2in,angle=0}
\vspace{-0.3in}
\caption{Plots of the predicted bands of $R_{AA}(p_{T})$ at $\sqrt{s}=62.4$ AGeV for both the ``squeezed" and ``uniform" distributions.  The figure on the left shows the predicted $R_{AA}$ for the ``squeezed" distribution while the figure on the right shows $R_{AA}$ for the ``uniform" distribution.}
\label{fig1}
\end{figure}
 The predicted $R_{AA}$ at 62.4 AGeV is obtained using the multiplicity systematics from PHOBOS \cite{Back:2001ae} that suggests the value $dN_{g}/dy(\sqrt{s}=62.4)\approx650-770$.  We expect from hadron gluon duality arguments that this multiplicity is reduced from about 1000 at 200 AGeV.  Therefore, we expect that $\bar{\epsilon}(62.4)\approx(0.65-0.77)\bar{\epsilon}(200)$.  The bands found by fitting to 200 AGeV PHENIX data can now be extrapolated to further bands at 62.4 AGeV.  The predicted $R_{AA}(p_{T})$ for both distributions can be seen in Fig. \ref{fig1}.  Using our multiplicity extrapolations we find that for the ``squeezed" distribution and $dN_{g}/dy(62.4)\approx 650(770)$, $0.42(0.50)<\bar{\epsilon}_{g}(62.4)<0.49(0.59)$.  Similarly, for the ``uniform" distribution and $dN_{g}/dy(62.4)\approx 650(770)$, $0.46(0.54)<\bar{\epsilon}_{g}(62.4)<0.52(0.62)$.  One of the things to note is that the ``uniform" distribution prediction is significantly flatter over $p_{T}$ than its ``squeezed" counterpart.  This is because once $\bar{\epsilon}>\frac{1}{2}$ the two distributions treat quenching very differently.  The ``uniform" distribution saturates to a uniform distribution over $0<\epsilon<1$ while the ``squeezed" distribution piles up closer to $\epsilon \rightarrow 1$ and causing a larger variation in the quench.  The calculations for the nuclear modification at $\sqrt{s}=62.4$ AGeV are consistent with Vitev \cite{Vitev:2004gn}.  Any deviations between the spectra can be attributed to the inclusion of models for Cronin interactions as well as initial parton $k_{T}$ smearing in  \cite{Vitev:2004gn} which are not included in the current calculations.
 
 \begin{figure}[htb]
 \hspace{1.5in}
\epsfig{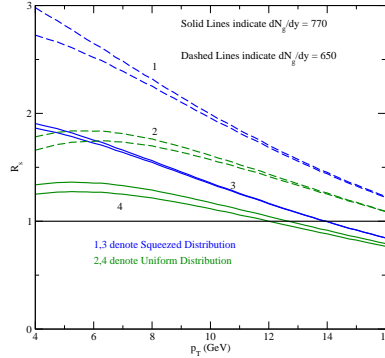}
\vspace{-0.2in}
\caption{The ratio for $R_{AA}$ at 62.4 AGeV to that at 200 AGeV.  Curves for ``squeezed" distributions are generally higher than curves for ``uniform" distributions.}
\label{fig2}
\end{figure}
Note that the predicted $R_{AA}(p_{T},\sqrt{s}=62.4)$ (Fig. \ref{fig1}) have a negative $p_{T}$ slope compared to the generally flat $R_{AA}$ at 200 AGeV.  This higher slope is caused by the earlier set in of the kinematic limits of the problem at lower energies.  The ``kinematic suppression" can be more robustly seen by calculating the observable $R_{s}(s)=\frac{R_{AA}(s)}{R_{AA}(200)}$ seen in Fig.  \ref{fig2}.  The $R_{s}$ curves have a distinct downwards slope due to the increasing power of the initial parton distributions.  $R_{s}$ is perhaps a better observable to use than $R_{AA}$ to observe the energy dependence of jet quenching as the uncertainty in the multiplicity extrapolations get canceled in the ratio.  $R_{s}$ also able to differentiate between the two types of fluctuation distributions (see Fig. \ref{fig2}).

\vfill\eject
\end{document}